\documentstyle[sprocl,epsf]{article}

\def\Journal#1#2#3#4{{#1} {\bf #2}, #3 (#4)}

\def\PRL{\em Phys. Rev. Lett.}
\def\PRD{{\em Phys. Rev.} D}

\def\beq{\begin{equation}}
\def\eeq{\end{equation}}
\def\bea{\begin{eqnarray}}
\def\eea{\end{eqnarray}}
\newcommand\bzeta{\mbox{\boldmath$\zeta$}}
\newcommand\bp{{\bf k}}

\newcommand\E{\epsilon}
\newcommand{\ds}[1]{#1 \hspace{-0.5em}/}  

\begin{document}

\title{
\hfill {\normalsize KUNS-1636 }\\
Magnetic instability of quark matter}

\author{Toshitaka Tatsumi}

\address{Department of Physics, Kyoto University, 
Kyoto 606-8502, JAPAN\\E-mail: tatsumi@ruby.scphys.kyoto-u.ac.jp} 


\maketitle\abstracts{ Spontaneous magnetization of quark liquid is
examined. It is pointed out that quark liquid has potential to be 
ferromagnetic at rather low densities.  
}

\section{Introduction}

Pulsars are rotating neutron stars emitting radio waves, X-rays or
gamma rays. Ordinary radio pulsars have a magnetic field of 
$O(10^{12 - 13})$G, which causes various radiations. The origin of such strong
magnetic field is still an open problem. Recently a new type of
neutron stars, called magnetars, has been proposed to explain the
observational data on pulsars, which should have an
extraordinary magnetic field of $O(10^{15})$ G \cite{ko}. There are
reported several magnetar candidates so far for anomalous X-ray
pulsars (AXP) and
pulsars associated with soft-gamma-ray repeaters (SGR). 

There has been a naive working hypothesis to understand the magnetic field
in neutron stars; if the magnetic flux of a main sequence star is
conserved during its evolution, the decrease in radius leads to an
increase in the magnetic field. For example, the sun, 
a typical main sequence star, 
has a magnetic field of $O(10^3)$G with the radius $R\sim
10^{10-11}$cm. By squeezing the radius to $10^6$cm for neutron stars
we have $O(10^{11-13})$G, which is consistent with observations for
radio pulsars. However, if this argument is extrapolated to explain the
intense of the magnetic field for magnetars, their radius should be 
$O(10^4)$cm, which is much less than the Schwartzschild radius of 
neutron stars with the canonical mass $M=1.4M_\odot$,
$R_{Sch}=2GM/c^2=4\times 10^5$cm. 

These observations
seems to enforce our reconsideration of the origin of the magnetic
field in neutron stars. Since there is a bulk hadronic matter 
beyond the nuclear density ($n_B\sim 0.16$fm$^{-3}$) inside neutron stars, 
it should be interesting to consider the hadronic 
origin of the magnetic field; ferromagnetism or spin-polarization of
hadronic matter may give such magnetic field. Unfortunately there has
been 
little suggestion about the possibility of spontaneous magnetization of
hadronic matter. We consider here
the possibility of ferromagnetism of quark liquid interacting with the 
one-gluon-exchange (OGE) interaction \cite{ta}.

One believes that there are deconfinement transition and  
chiral symmetry restoration at several times the nuclear  density, 
while their critical
densities have not been fixed yet. One interesting suggestion is that
three-flavor symmetric quark matter (strange quark matter) around 
or above the nuclear density may be the true ground state of
matter \cite{ch}. If this is the case, 
strange quark stars, where quarks occupy almost whole the inner 
region of stars,  can exist in a different branch
from the neutron-star branch in the mass-radius plane. 
Otherwise quark matter may exist in the
small core region of neutron stars. We shall see our results should
give an origin of the strong magnetic field 
in the context of strange quark-star scenario.

\section{Ferromagnetism of quark liquid}

Quark liquid should be totally color singlet (neutral), which means
that only the exchange interaction between quarks is relevant there.
This may remind us of electron system with the Coulomb interaction
in a neutralizing positive charge background.
In 1929 Bloch first suggested a possibility of ferromagnetism of
electron system \cite{bl}. He has shown that there is a trade off between the
kinetic and the exchange energies as a function of density, 
the latter of which favors the spin 
alignment due to the Pauli principle. This was a beginning of of the
concept of itinerant magnetism.
In the following we discuss the
possibility of ferromagnetism of quark liquid on the analogy with
electron gas. 

It is to be noted that there is one big difference
between them; quarks should be treated in a relativistic way. The
concept of the direction of spins is not well defined in relativistic
theories, while each quark has two polarization degrees of
freedom. Here we define the spin-up and -down states in the rest frame of
each quark. Then the projector onto states of definite polarization is 
given by 
\beq
P(a)=\frac{1}{2}(1+\gamma_5\ds{a})
\eeq
with the 4-peudovector $a$,
\beq
{\bf a}=\bzeta+\frac{\bp(\bzeta\cdot\bp)}{m_q(E_k+m_q)}, 
~a^0=\frac{\bzeta\cdot\bp}{m_q}
\label{aa}
\eeq
for a quark moving with the momentum $k=(E_k,\bp)$ \cite{la}. The 4-peudovector
$a$ is reduced into the axial vector $\bzeta$ ($|\bzeta|=1$) 
in the rest frame, which 
is twice the mean spin vector in the rest frame. Actually if we choose 
$\bzeta$ along the $z$ axis, $\bzeta=(0,0,\pm 1)$, we can see each
value corresponds to the spin-up or -down state. The mean value of the 
spin is given by 
\beq
\bar{\bf s}=\frac{1}{2}\frac{m_q}{E_k}\left(\bzeta
+\frac{\bp(\bzeta\cdot\bp)}{m_q(E_k+m_q)}\right).
\label{ab}
\eeq
Finally the  projection operator $P(a)$ gives the polarization density 
matrix $\rho$,
\beq
\rho(k, \zeta)
=\frac{1}{2m_q}(\ds{k}+m_q)P(a), \quad P(a)=\frac{1}{2}(1+\gamma_5\ds{a}).
\label{ac}
\eeq

The exchange interaction between two quarks with momenta ${\bf k}$ 
and ${\bf q}$ is given by
\beq
 f_{{\bf k}\zeta,{\bf q}\zeta'}
=\frac{m_q}{E_k}\frac{m_q}{E_q}{\cal M}_{{\bf k}\zeta,{\bf q}\zeta'}.
\label{ad}
\eeq
${\cal M}_{{\bf k}\zeta,{\bf q}\zeta'}$ is the usual Lorentz
invariant matrix element, and is evaluated with the help of the
polarization density matrix (\ref{ac})
\bea
{\cal M}_{{\bf k}\zeta,{\bf q}\zeta'}
&=&g^2\frac{1}{9}{\rm tr}(\lambda_a/2\lambda_a/2)\frac{1}{4}{\rm tr}
\left[\gamma_\mu\rho(k,\zeta)\gamma^\mu\rho(q,\zeta')\right]
\frac{1}{(k-q)^2}\\
&=&g^2\frac{2}{9}[2m_q^2-k\cdot q-m_q^2 a\cdot b]\frac{1}{(k-q)^2},
\label{ae}
\eea
where the 4-pseudovector $b$ is given by the same form as in Eq.~(\ref{aa}) for
the momentum ${\bf q}$. 

Although the vector $\bzeta$ of each quark may point in a different direction
on the two dimentional sphere $S^2$, we assume here it along the same
direction, say $z$ axis. 
The exchange energy for quark liquid is then given by the integration of
the interaction (\ref{ad}) over the two Fermi seas with the spin-up
and -down states; eventually, it consists of two contributions,
\beq
\epsilon_{ex}=\epsilon_{ex}^{non-flip}+\epsilon_{ex}^{flip}.
\eeq
The first one arises from the interaction between quarks with the
same polarization, while the second one with the opposite polarization.
The non-flip contribution is the similar one as in electron gas, while 
the flip contribution is a genuine relativistic effect and never
appears in electron gas. We shall see that this relativistic effect
leads to a novel mechanism of ferromagnetism of quark liquid.

\section{Examples}

We show some results about the total energy of quark liquid, 
$\E_{tot}=\E_{kin}+\E_{ex}$, by adding the kinetic term $\E_{kin}$. Since
gluons have not the flavor quantum numbers, we can consider one flavor 
quark matter without loss of generality. Then quark number density 
directly corresponds to baryon number density, if we assume the three flavor 
symmetric quark matter as mentioned in \S 1.

\begin{figure}[t]
\centerline{
\epsfysize=8cm
\epsfbox{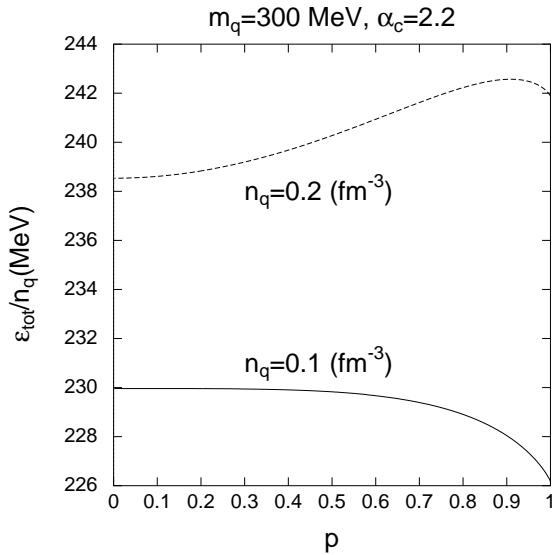}
}
\caption{Total energy of quark liquid as a function of the
polarization parameter for densities $n_q=0.1, 0.2$fm$^{-3}$.
}
\end{figure}

There are two parameters in our theory: the quark mass $m_q$ and the
quark-gluon coupling constant $\alpha_c$. These values are not well
determined so far. In particular, the value of quark mass involves subtle
issues; it depends on the current or constituent quark picture and 
may be also related to the existence of chiral phase transition. 
Here we allow some range for these parameters and
take, for example,  a fiducial set, $m_q=300$MeV for strange quark 
and $\alpha_c=2.2$, given by the MIT bag model \cite{de}. In Fig.1 two
results are presented as functions of the polarization parameter $p$
defined by the difference of the number of the spin-up and -down
quarks, $n_q^+-n_q^-\equiv pn_q$.
The results clearly show
that the ground state should be ferromagnetic for lower density, while
it is in the paramagnetic phase for higher density. The phase
transition is of first order and its critical
density is around $n_q^c\simeq 0.16$fm$^{-3}$ in this case, 
which corresponds to the
nuclear density for flavor symmetric quark matter. Note that there is
a metastable ferromagnetic state (the local minimum) 
even above the critical density. This ferromagnetic phase is a
spontaneously symmetry broken state with respect to the rotational
symmetry: the order parameter is the mean value of 
$\bzeta$, $\langle\bzeta\rangle$, and symmetry is broken from
$G=O(3)$ to $H=O(2)$ once $\langle\bzeta\rangle$ takes a special 
direction on $S^2$.


Magnetic properties of quark liquid 
are characterized by three quantities, $\delta \E,
\chi$ and $\eta$;
$\delta \E\equiv \E_{tot}(p=1)-\E_{tot}(p=0)$, which 
is the measure for ferromagnetism to appear in the ground state.
For small $p\ll 1$,
\beq
\E_{tot}-\E_{tot}(p=0)=\chi^{-1} p^2+O(p^4).
\label{ha}
\eeq
$\chi$ is 
proportional to the magnetic susceptibility and plays an important
role if the phase transition is of second order. In our case it is
less relevant since the phase transition is of first order. Finally, 
$\eta\equiv\partial \E_{tot}/\partial p~|_{p=1}$,
which is the measure for metastability to to exist. In Fig.2 the density
dependence of three quantities are given for a fiducial set of parameters.
We can see that ferromagnetic phase is the ground state below $n_q^c$, 
while the metastable state is possible up to rather high densities. 

\begin{figure}[t]
\centerline{
\epsfysize=8cm
\epsfbox{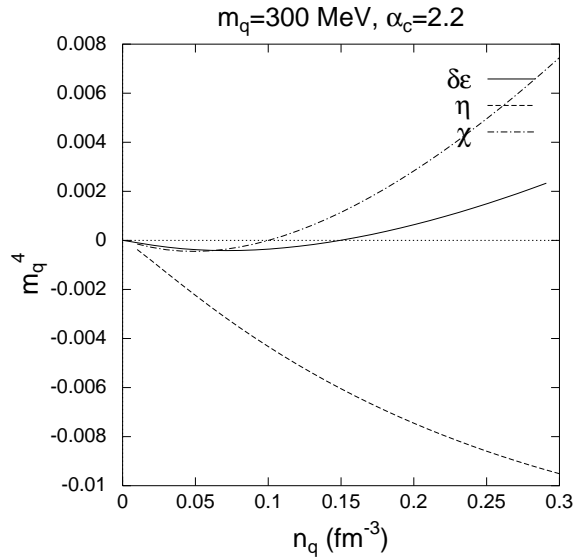}
}
\caption{Density dependence of three quantities. The negative regions
for $\delta\E$ and $\eta$ indicate the ferromagnetic phase and the
existence of the metastable ferromagnetic state.
}
\end{figure}


Finally we present a phase diagram in the $m_q - \alpha_c$ plane for
$n_q=0.3$fm$^{-3}$, which corresponds to about twice the nuclear
density for flavor symmetric quark matter. The region above the solid 
line shows the ferromagnetic phase and that above the
dashed line indicates the existence of the matastable state. For 
massive quarks with the
large mass, which may correspond to the current $s$ quarks
or the constituent quarks before chiral symmetry restoration, 
the ferromagnetic state is favored for small coupling 
constant due to the same mechanism as in electron gas. The
ferromagnetic state is favored again for light quarks with small mass, 
which may
correspond to the current $u, d$ quarks, while the nonrelativistic 
calculation does not show such tendency. Hence this is due to a 
genuine relativistic effect, where the spin-flip interaction plays an
essential role. 

\begin{figure}[t]
\centerline{
\epsfysize=8cm
\epsfbox{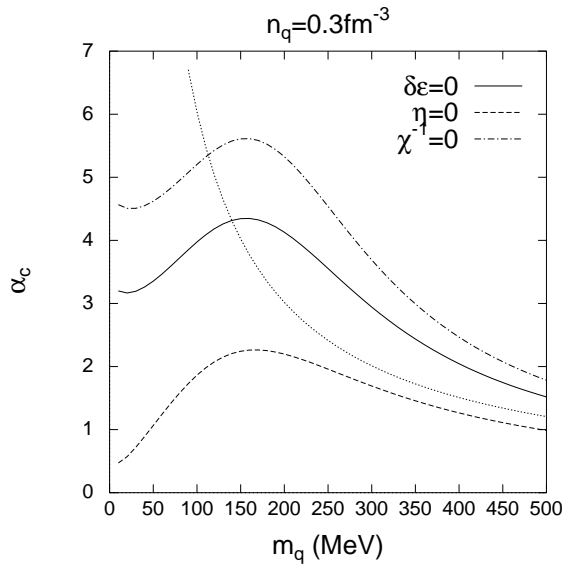}
}
\caption{Phase diagram in the mass ($m_q$)- the coupling constant
($\alpha_c$) plane. $\delta\E$ in the nonrelativistic
calculation is depicted for comparison (the dotted line) .
}
\end{figure}


\section{Summary and Concluding remarks}

We have seen that the ferromagnetic phase 
is realized at low densities 
and the metastable state is plausible up to rather high densities for
a reasonable range of the QCD parameters. 
If a ferromagnetic quark liquid exists stably or metastably around or
above nuclear density, it has
some implications on the properties of strange quark stars and strange
quark nuggets. They should be magnetized in a macroscopic
scale. 
For quark stars with the quark core of $r_q$, simply assuming 
the dipolar magnetic field, we can estimate its strength  at the
surface $R$, 
\beq
B_{max}=\frac{8\pi}{3}\left(\frac{r_q}{R}\right)^3\mu_qn_q,
\label{gc}
\eeq
amounts to order of $O(10^{15-17})$G for $r_q\sim O(R)$ and 
$n_q=O(0.1)$fm$^{-3}$ 
, which should be large enough for magnetars,
using the quark 
magnetic moment $\mu_q\sim
\mu_N$($\mu_N: $nuclear magneton$\sim 5\times 10^{-24}{\rm erg}\cdot
{\rm gauss}^{-1}$) for massive quarks and $10^2\mu_N$ for light
quarks. Hence it might be interesting to model SGR or AXP using our idea. 

We have found that 
ferromagnetic instability is feasible not only
in the massive quark system but also in the light quark system: 
the spin-nonflip contribution is dominant in the nonrelativistic case 
as in electron gas, while a novel mechanism appears  
as a result of the large spin-flip contribution in the relativistic case.

Our calculation is basically a perturbative one and the Fermi sea 
remains in a spherical shape. However, if we get more insight about the
ferromagnetic phase, we must solve the Hartree-Fock equation and
thereby derive a self-consistent mean-field for quark
liquid. Moreover, we need to examine the long range correlation among
quarks by looking into the ring diagrams, which has been known to
be important in the calculation of the susceptibility of electron gas.

\end{document}